\documentclass[fleqn,twoside]{article}
\usepackage{espcrc2}
\usepackage{graphicx}
\usepackage{epsfig}
\newcommand{\AmS}{{\protect\the\textfont2
  A\kern-.1667em\lower.5ex\hbox{M}\kern-.125emS}}

\title{Multiparticle amplitudes at one-loop: an algebraic/numeric approach}

\author{T. Binoth\address{Institut f\"ur Theoretische Physik, 
        Universit\"at W\"urzburg, \\ 
        Am Hubland, D-97074 Würzburg, Germany}%
        \thanks{This work was supported by the Bundesministerium f\"ur
Bildung und Forschung (BMBF, Bonn, Germany) under the contract
number 05HT1WWA2.}}
       
\begin{document}

\begin{abstract}
We discuss algebraic/numeric methods to compute one-loop corrections
for multiparticle/jet production cross sections. By using efficient reduction algorithms
a compact expression for the $ggg\gamma\gamma \to 0$ amplitude is obtained.
Further a numerical approach for 6-point 1-loop diagrams is presented.  
\vspace{1pc}
\end{abstract}

% typeset front matter (including abstract)
\maketitle

\section{INTRODUCTION}

\begin{figure}[htb]
\unitlength=1mm
\begin{picture}(80,25)
\put(10,-5){\epsfig{file = 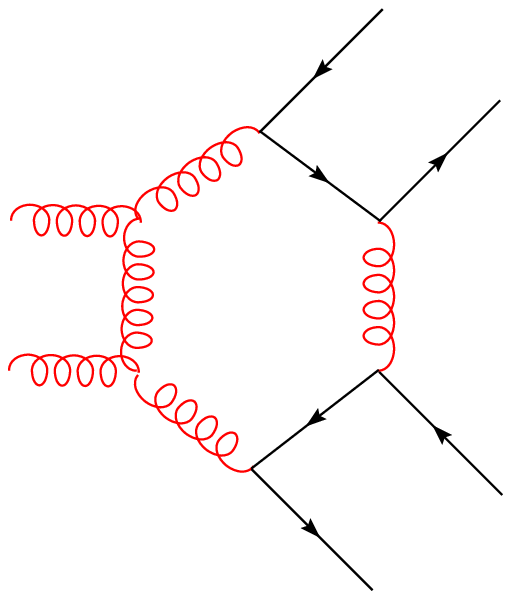,height=3.0cm}}
%\put(16,5){$g$}
%\put(16,13){$g$}
%\put(40,-5){$b$}
%\put(47,0){$\bar b$}
%\put(40,25){$b$}
%\put(47,20){$\bar b$}
%\put(10,-5){\epsfig{file = fighex_yygg.eps,height=2.5cm}}
%\put(6,5){$g$}
%\put(6,13){$g$}
%\put(30,-3){$\gamma$}
%\put(37,0){$g$}
%\put(30,25){$\gamma$}
%\put(37,20){$g$}
\put(40,-5){\epsfig{file = 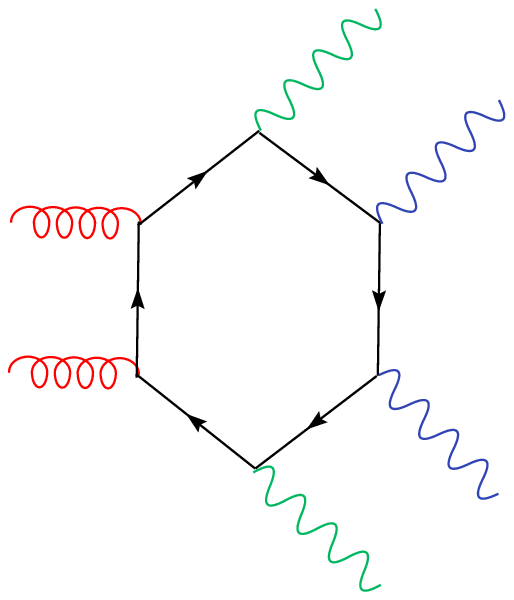,height=3.0cm}}
%\put(86,5){$g$}
%\put(86,13){$g$}
%\put(110,-5){$\gamma$}
%\put(117,0){$Z$}
%\put(110,25){$\gamma$}
%\put(117,20){$Z$}
%\put(102,14){$t$}
\end{picture}
\caption{Typical Hexagon graph for the 1-loop amplitudes $gg\to b\bar b b \bar b$ and  $gg\to ZZ\gamma\gamma$.\label{FigHex}}
\end{figure}

The theoretical description of multi-particle production at the one-loop level
is a very challenging task, as  the  complexity of 
the Feynman diagrammatic approach grows exponentially with the number of external partons. 
%Whereas the calculation of partonic $2 \to 2$ amplitudes at one-loop
%is meanwhile standard, already the number of known $2 \to 3$  1-loop amplitudes is very restricted.
%Notable exceptions relevant for multi-jet \cite{Bern:1993mq,Kunszt:1994tq,Bern:1997sc,Yasui:2002bn} 
%and Higgs physics exist 
%\cite{Reina:2001bc,Beenakker:2002nc,Beenakker:2001rj,Belanger:2003ya,Belanger:2002ik,Belanger:2003nm}. 
No Standard Model process which has generic  $2\to 4$ 
kinematics is computed at the one-loop level although 
this is highly relevant
for  many Higgs boson search channels  at the LHC, like gluon fusion
and weak boson fusion, where additional jets have to be tagged to improve the
signal to background ratio. For signal reactions like 
$PP \to H + 0,1,2$ jets, with $ H \to \gamma\gamma,WW^*,\tau^+\tau^-$
which are available at one-loop level, many backgrounds remain to be calculated. 
As an example for needed calculations consider 
$PP \to b\bar b b\bar b + X$, $PP \to \gamma\gamma + 2\;\mbox{jets} + X$ or 
$PP \to ZZ + \gamma\gamma + X$, which require the evaluation of hexagon graphs like the ones 
given in Fig.~\ref{FigHex}. 

Te computation of the related amplitudes relies on efficient 
 methods for the evaluation of the corresponding Feynman graphs.  
In the next section we shortly review  our reduction formalism. 
As an example for the efficiency of our methods 
we discuss the 5-point 1-loop amplitude $gg \to \gamma\gamma g$
in Section 3. 
It seems to be feasible to apply the presented techniques 
also for 6-point processes, as long as the internal masses
of the problem can be neglected.  In  \cite{Binoth:2002qh,Binoth:2001vm}
we have shown that in the massless  Yukawa model our formalism leads to
compact expressions.       
Going to the massive case leads generally to much more involved
expressions and in that case numerical methods seem to be preferable.
An approach for the numerical evaluation of 6-point Feynman diagrams
is outlined in Section 4.

\section{REDUCTION FORMALISM}\label{sec:redu}
In the Feynman diagrammatic approach any one-loop amplitude can be represented as
a linear combination of factors which contain the group theoretical information  
and tensor one-loop integrals:
\begin{eqnarray*}
\Gamma^{\{ c \},\{\lambda \}}({p_j,m_j}) = \sum\limits_{\{ c_i\}} f^{\{c_i\}}{\cal G}^{\{\lambda\}}
\end{eqnarray*}
where
\begin{eqnarray*}
{\cal G}^{\{\lambda\}} &=& \int\frac{d^nk}{i\pi^{n/2}}\frac{ {\cal N}^{\{\lambda\}} }{ (q_1^2-m_1^2)\dots(q_N^2-m_N^2) }\\
&=& \sum\limits_{R}  {\cal N}^{\{\lambda\}}_{\mu_1,\dots,\mu_R}\; I_N^{\mu_1\dots\mu_R}({p_j,m_j}) \\
I_N^{\mu_1\dots\mu_R}&=& \int\frac{d^nk}{i\pi^{n/2}}\frac{ k^\mu_1 \dots k^\mu_R }{ (q_1^2-m_1^2)\dots(q_N^2-m_N^2) }
\end{eqnarray*}
The propagator momenta are defined by $q_j = k-r_j=k-p_1\dots -p_j$.
To separate the Lorentz structure from the integrals it is useful to express the
tensor integrals in terms of scalar integrals with nontrivial numerators. % \cite{}.
\begin{eqnarray*}
&&I^{\mu_1 \dots \mu_R}_N = \sum \limits_{m=0}^{[R/2]} 
\left(-\frac{1}{2}\right)^m \\&&
\sum \limits_{j_1,\dots,j_{R-2m}=1}^{N-1} 
\left[ g^{\cdot\cdot}_{(m)} r_{j_1}^{\cdot}\dots r_{j_{R-2m}}^{\cdot} \right]^{\{\mu_1 \dots \mu_R\}}\\&&
I_N^{n+2m}(j_1,\dots,j_{R-2m})
\end{eqnarray*}
$[R/2]$ is the smaller nearest integer to $R/2$.  
The bracket with the Lorentz indices as superscripts stands for the sum over all distinguishable
distributions of the Lorentz indices to the metric tensors  and external momenta.
The separation of the kinematical information allows  to sort the amplitude
into gauge invariant subsets.  
Basic ingredient of the given formula is the  Feynman parameter integral defined in $D=n+2m$ dimensions:
\begin{eqnarray}\label{basic_integral}
&&I_N^D(j_1,\dots ,j_R) = (-1)^N \Gamma (N-D/2)  \nonumber\\&&
\int_{0}^{\infty} d^Nx \,\delta(1-\sum\limits_{l=1}^N x_{l})
\frac{x_{j_1}\dots x_{j_R}}{(x\cdot S\cdot x -i\delta)^{N-D/2}} \nonumber\\&&
\end{eqnarray}
In \cite{Binoth:1999sp} we have derived a reduction formula for such
parameter integrals for general $N,R$ in arbitrary dimensions $D$.
It is based on differentiation by parts in parameter space.
The derived formula maps rank $R$ $N$-point integrals in $D$ dimensions
to rank $R-1$ $N$-point integrals and higher dimensional integrals
with lower rank. As the latter are IR finite, a separation of
IR divergent and IR finite terms can be obtained in this way
which defines an approach for a semi-numeric method.
After extraction of all UV/IR poles the remaining integrals can
be treated numerically. If one wants to proceed analytically  
one has to iterate the formula.
In this way one can show that arbitrary $N$-point
Feynman integrals can be expressed in terms of $n=4-2\epsilon$ dimensional
bubble and triangle functions and  $n+2$ dimensional boxes.
More details on reduction formalisms 
can be found in \cite{Binoth:1999sp,Bern:1994kr}.

\section{THE LOOP AMPLITUDE $gg\to \gamma\gamma g$}
To give an example for our algebraic approach 
we have considered
the 5-point 1-loop  amplitude $gg\to \gamma\gamma g$ \cite{Binoth:2003xk}. 
This amplitude is indirectly known from the 1-loop
5-gluon amplitude \cite{Bern:1993mq} 
by turning gluons into photons.

We define all particles as incoming.
\begin{eqnarray}
&& \gamma(p_1,\lambda_1) + \gamma(p_2,\lambda_2) 
+ g( p_3,\lambda_3,c_3 ) \nonumber \\
&& + g( p_4,\lambda_4,c_4 ) + g( p_5,\lambda_5,c_5 ) \to 0
\end{eqnarray} 
In hadronic collisions this amplitude is relevant for the 
production of photon pairs
in association with a jet and as such a contribution of the background
to the Higgs boson search channel $H\to\gamma\gamma + \mbox{jet}$.
For a phenomenological analysis see \cite{deFlorian:1999tp,DelDuca:2003uz}.
The colour  structure of this amplitude  can be written  as
\begin{eqnarray*}
\Gamma^{\{\lambda_j,c_j\}}[\gamma\gamma g g g \to 0] 
= \frac{Q_q^2 g_s^3}{i \pi^2} f^{c_3c_4c_5} {\cal A}^{\lambda_1\lambda_2\lambda_3\lambda_4\lambda_5}
\end{eqnarray*}
${\cal A}^{\lambda_1\lambda_2\lambda_3\lambda_4\lambda_5}$ are helicity dependent
linear combinations of scalar integrals and a constant term which is a remnant of
two-point functions with coefficients of order $(D-4)$.
Six independent helicity components exist: +++++,++++\,--,\\--\,++++,--\,--\,+++,
+++\,--\,--,\,--\,+++\,--.
As the amplitude is finite one expects that all 3-point functions which carry spurious
infrared poles cancel. The function basis of the problem is thus reduced to 2-point functions 
\begin{eqnarray*}
I_2^D(s_{ij}) = \frac{\Gamma(1+\epsilon)\Gamma(1-\epsilon)^2}{\Gamma(2-2\epsilon)} \frac{(-s_{ij})^{-\epsilon}}{\epsilon} ,
\end{eqnarray*}
4-point functions in 6 dimensions  written as \cite{Binoth:2001vm}
\begin{eqnarray*}
F_1(s_{j_1j_2},s_{j_2j_3},s_{j_4j_5}) = \frac{I_4^6(p_{j_1},p_{j_2},p_{j_3},p_{j_4}+p_{j_5})}{s_{j_1j_3}}\nonumber
\end{eqnarray*}
and constant terms. 
From unitarity one expects that the  +++++\,,\,++++\,--\,,\,--\,++++ amplitudes should be polynomial.
The other helicity amplitudes will also contain non-polynomial functions
like logarithms and dilogarithms contained in  $I_2^D$ and $F_1$.

To give an example for a compact helicity amplitude we show here the result for 
${\cal A}^{--+++}$ only. The remaining ones which
have also compact representations can be found in \cite{Binoth:2003xk}.
The result is expressed in terms of field strength 
tensors ${\cal F}^{\mu\nu}_j=p^\mu_j\epsilon^\nu_j-p^\nu_j\epsilon^\mu_j$
where $\epsilon_j^\pm$ are the polarization vectors of the gluons and photons.

We split the result of ${\cal A}^{--+++}$ into three pieces with indices $F,B,1$, which belong to the  
 part proportional to 6-dimensional 
boxes $F_1$,  a part containing bubble graphs $I_2^D$,  and a constant term, respectively.   
\begin{eqnarray}
{\cal A}^{--+++} = {\cal A}^{--+++}_F + {\cal A}^{--+++}_B + {\cal A}^{--+++}_1 \nonumber
\end{eqnarray} 
We find
\begin{eqnarray}
{\cal A}^{--+++}_F = \frac{\mbox{Tr}({\cal F}_1^-{\cal F}_2^-)\mbox{Tr}({\cal F}_3^+{\cal F}_4^+)}{s_{12}^2 s_{34}^2} \hspace{2cm}\nonumber\\ 
\Bigl[ C^{--+++}_F\; p_1\cdot {\cal F}_5^+\cdot p_3 - (  3 \leftrightarrow 4 ) \Bigr] F_1(s_{13},s_{14},s_{25})\nonumber\\ 
- (  4 \leftrightarrow 5 ) - (  5 \leftrightarrow 3 ) + (  1 \leftrightarrow 2 )  \nonumber\\ 
       - (  1 \leftrightarrow 2, 4 \leftrightarrow 5 )
       - (  1 \leftrightarrow 2, 5 \leftrightarrow 3 )\nonumber\\
{\cal A}^{--+++}_B = \frac{\mbox{Tr}({\cal F}_1^-{\cal F}_2^-)\mbox{Tr}({\cal F}_3^+{\cal F}_4^+)}{s_{12}^2 s_{34}^2} \hspace{2cm}\nonumber\\
\Bigl[ C^{--+++}_B\; p_1\cdot {\cal F}_5^+\cdot p_3 - (  3 \leftrightarrow 4 ) \Bigr] I_2^D(s_{15})\nonumber\\ 
- (  4 \leftrightarrow 5 ) - (  5 \leftrightarrow 3 ) + (  1 \leftrightarrow 2 )  \nonumber\\
       - (  1 \leftrightarrow 2, 4 \leftrightarrow 5 ) - (  1 \leftrightarrow 2, 5 \leftrightarrow 3 )\nonumber\\      
{\cal A}^{--+++}_1 = \frac{\mbox{Tr}({\cal F}_1^-{\cal F}_2^-)\mbox{Tr}({\cal F}_3^+{\cal F}_4^+{\cal F}_5^+)}{s_{34}s_{45}s_{35}}
\hspace{1.5cm}\nonumber
\end{eqnarray} 
The indicated permutations have to be applied to the  indices of the field strength tensors, 
momenta and  Mandelstam variables. 
The coefficients are
\begin{eqnarray}
C^{--+++}_F = \frac{1}{2}\frac{s_{12}^2-2s_{13}s_{14}}{s_{35}s_{15}} 
                         - \frac{s_{14}}{s_{34}}- \frac{s_{14}}{s_{35}}\hspace{.5cm}\nonumber\\
C^{--+++}_B = \frac{s_{45}}{s_{15}}\left[\frac{s_{13}+s_{35}}{s_{14}+s_{45}}+\frac{s_{14}+s_{45}}{s_{13}+s_{35}}\right]\hspace{.5cm}
\nonumber\\
                         +\frac{s_{45}^2s_{13}}{s_{15} s_{35}^2}
			 +\frac{s_{14}s_{35}}{s_{15} s_{45}}+2\frac{(s_{15}+s_{45})^2}{s_{35}^2}
			 \nonumber\\ 
			-\frac{s_{14}s_{45}}{s_{15} s_{35}}
			+\frac{s_{14}+s_{24}}{s_{45}}
			+2\frac{s_{14}(s_{15}+s_{45})}{s_{35}^2}
			\nonumber\\ 
			+\frac{s_{12}-s_{14}-s_{35}}{s_{14}+s_{45}}
			+\frac{s_{23}^2s_{15}}{s_{35}^2(s_{13}+s_{35})}
			+\frac{2s_{45}+s_{15}}{s_{13}+s_{35}}\nonumber\\ 
			-2\frac{(s_{15}+s_{45})s_{23}}{s_{35}(s_{13}+s_{35})}
			-\frac{(2s_{45}+s_{15})}{s_{35}}\nonumber\\ 
			+\frac{s_{13}(2s_{45}+s_{15})}{s_{35}^2}-\frac{s_{13}+s_{35}}{s_{15}}-\frac{s_{45}^2}{s_{35} s_{15}}
			\nonumber
\end{eqnarray} 
In the given expressions the  $S_2 \otimes S_3$ symmetry 
under exchange of the two photons and the three gluons is manifest 
after taking into account the omitted colour factor.

The result indicates that with our approach indeed a compact representation
of complicated loop amplitudes can be obtained. The application of our approach to
relevant 6-point amplitudes is presently under study. 

\section{NUMERICAL APPROACH}

Due to the complexity of the analytic approach if massive particles 
are present, a numerical approach seems to be more appropriate 
to tackle different types of one-loop amplitudes  in a unified
and efficient way. 

Recently a great activity in that direction with many new ideas
can be observed 
\cite{Nagy:2003qn,Giele:2004iy,delAguila:2004nf,Ferroglia:2002mz}.

\subsection{Reduction to basic building blocks}

As basic building blocks for an amplitude in our numeric approach, 
we  choose scalar 2-point functions $I^n_2$ and 
3-point functions $I^n_3$  and $n+2$ dimensional 
box functions $I^{n+2}_4$ with nontrivial numerators.  
The latter are infrared finite. 
Possible UV singularities are only contained in the 2-point 
functions and their subtraction is straightforward. 
The (soft and collinear) IR singularities are, as a result of the 
reduction, only contained in 2-point functions and 3-point functions
with one or two light-like legs. In this form, 
they are easy to isolate and to subtract from the amplitude. 
After reduction and separation of the divergent parts, 
we are left with finite integrals $I^n_3(j_1,j_2,j_3)$ and 
$I^{n+2}_4(j_1,j_2,j_3,j_4)$, with nontrivial 
numerators. As numerical stability  problems are entirely from the denominators
we discuss only the  case of scalar integrals with trivial numerators here.
Systematic methods for the combination 
of the IR divergences from the virtual corrections with 
their counterparts from the real emission contribution already exist
(\cite{Catani:2002hc} and references therein).

In this section we focus on the evaluation of a finite 6 point scalar 
integral. As a first step we reduce the hexagon integral to 
box and triangle functions which are the basic building blocks of 
the reduction.

\subsection{Parameter representation of basic building blocks}

To evaluate the box and triangle functions numerically,  
we first  perform a sector decomposition.
\begin{eqnarray}
\label{sectordeco}
1 &=& \Theta(x_1>x_2,\dots,x_N) +\Theta(x_2>x_1,\dots,x_N) \nonumber\\
&& + \dots + \Theta(x_N>x_1,\dots,x_{N-1})
\end{eqnarray}
for the integration over $N$ parameters ($N=3$ for the triangle, $N=4$ 
for the box).
The step function $\Theta$ is defined as 1 if the inequality of its argument is 
fulfilled, and 0 else.
Now, we carry out  one parameter integration explicitly. We show the explicit 
expressions only for the triangle integral, 
the ones for the box are analogous and can be found in \cite{Binoth:2002xh}. 
We obtain
\begin{eqnarray}
I_3^{D}(s_1,s_2,s_3,m_1^2,m_2^2,m_3^2) = \hspace{1cm}\nonumber\\\hspace{1cm}
\Bigl[  
        S_{Tri}^D(s_2,s_3,s_1,m_2^2,m_3^2,m_1^2) \nonumber\\\hspace{1cm}
       +S_{Tri}^D(s_3,s_1,s_2,m_3^2,m_1^2,m_2^2) \nonumber\\\hspace{1cm}
      + S_{Tri}^D(s_1,s_2,s_3,m_1^2,m_2^2,m_3^2) \Bigr]\nonumber
\end{eqnarray}
with
\begin{eqnarray}\label{stri}
S_{Tri}^{D=4}(s_1,s_2,s_3,m_1^2,m_2^2,m_3^2) \hspace{2cm}\nonumber\\
=\int\limits_0^1 dt_1dt_2 
\frac{1}{(1+t_{1}+t_2)}\frac{1}{A t_2^2 + B t_2 + C - i \delta}\nonumber\\
\end{eqnarray}
\begin{eqnarray}
&&A = m_2^2 \nonumber\\
&&B = (m_1^2+m_2^2-s_2) t_1 + m_2^2 + m_3^2 -s_3 \nonumber\\
&&C = m_1^2 t_1^2  + ( m_1^2+m_3^2-s_1) t_1 + m_3^2 \nonumber\\
&&R = B^2 - 4 A C + i \delta \;\nonumber\\
&&T = 2 A ( 1+t_1 ) - B\nonumber
\end{eqnarray}
The discussion is also valid in the case of vanishing masses or invariants, 
as long as the functions remain IR finite. Note that if infrared divergences
are present the triangle integrals can typically be treated analytically.
The $(n+2)$-dimensional box function are infrared finite for any 
physically relevant kinematics.

\subsection{Singularity structure}
Starting from (\ref{stri}) one integration is performed explicitly.
In order to analyse the singularity structure of the
integrands, we then separate imaginary and real part.
One obtains
\begin{eqnarray}
S_{Tri}^{D=4}(s_1,s_2,s_3,m_1^2,m_2^2,m_3^2) \hspace{2.5cm}\nonumber\\
= \int\limits_0^1 dt_1  \frac{4 A}{T^2-R} 
\Bigl\{
\Bigl[
\log(2A+B+T) - \log(B+T)
\Bigr] \nonumber\\
+\Theta(R<0)
\Bigl[ 
       \frac{\log(C) - \log(A+B+C)}{2}  \nonumber\\
+\frac{T}{\sqrt{-R}}
\Bigl(  
   \arctan\left(  \frac{\sqrt{-R}}{B} \right) 
  -\arctan\left(  \frac{\sqrt{-R}}{2 A+B}\right) \nonumber\\
  + \pi\;\Theta(B < 0 < 2A+B)  
\Bigr)  
\Bigr] \nonumber\\
+\Theta(R>0)
\Bigl[ 
\frac{T - \sqrt{R}}{2\,\sqrt{R}}
\Bigl(
 \log\left( | 2 A + B - \sqrt{R} |\right) \nonumber\\
-\log\left( | B - \sqrt{R} |\right) -i \pi \Theta( B < \sqrt{R}< 2A+B)
\Bigr)\nonumber\\
- \frac{T + \sqrt{R}}{2\,\sqrt{R}}
\Bigl(
 \log\left( | 2 A + B + \sqrt{R} |\right) \nonumber\\
-\log\left( | B + \sqrt{R} |\right) +i \pi \Theta( B < -\sqrt{R}< 2A+B)
\Bigr)
\Bigr]
\Bigr\}\nonumber\label{Stri_final}
\end{eqnarray} 
Three regions which lead to an imaginary part can be distinguished:
\begin{description}
\item[Region I:] $A+B+C>0, -2A<B<0$,\\ $C>0 \Leftrightarrow 
                 ( B<\pm \sqrt{R}<2A+B )$.
\item[Region II:] $A+B+C>0, C<0 \Leftrightarrow 
               (B< \sqrt{R}<2A+B)\, \mbox{and not} \,(B<-\sqrt{R}<2A+B)$.
\item[Region III:]  $A+B+C<0, C>0 \Leftrightarrow$ \\ 
               $(B<-\sqrt{R}<2A+B)$ and not $\,(B<\sqrt{R}<2A+B)$.
\end{description}  
Region I is an overlap region where the imaginary part 
has two contributions. In regions II and III only one of the 
$\Theta$--functions  contributes.
Note that  the 
box function $I_4^{D=6}$ has the same singularity structure \cite{Binoth:2002xh}. 
As $I^{D=4}_3$ and $I_4^{D=6}$ are the basic building blocks, 
this analysis of the singularity structure is done once and for all. 
Knowing the critical region of integration it is possible to map
out the singularities by adequate parameter transformations.

\subsection{Numerical integration}

To demonstrate the practicality of our method to evaluate
multi-leg integrals, we show in
Fig.~\ref{hexagon-scan} a scan of the $2m_t = 350$ GeV
threshold of the 4-dimensional scalar hexagon function for a realistic kinematical
configuration.
\unitlength=1mm
\begin{figure}[h]
\begin{picture}(80,83)
\put(10, 40){\includegraphics[width=5.cm, angle=90]{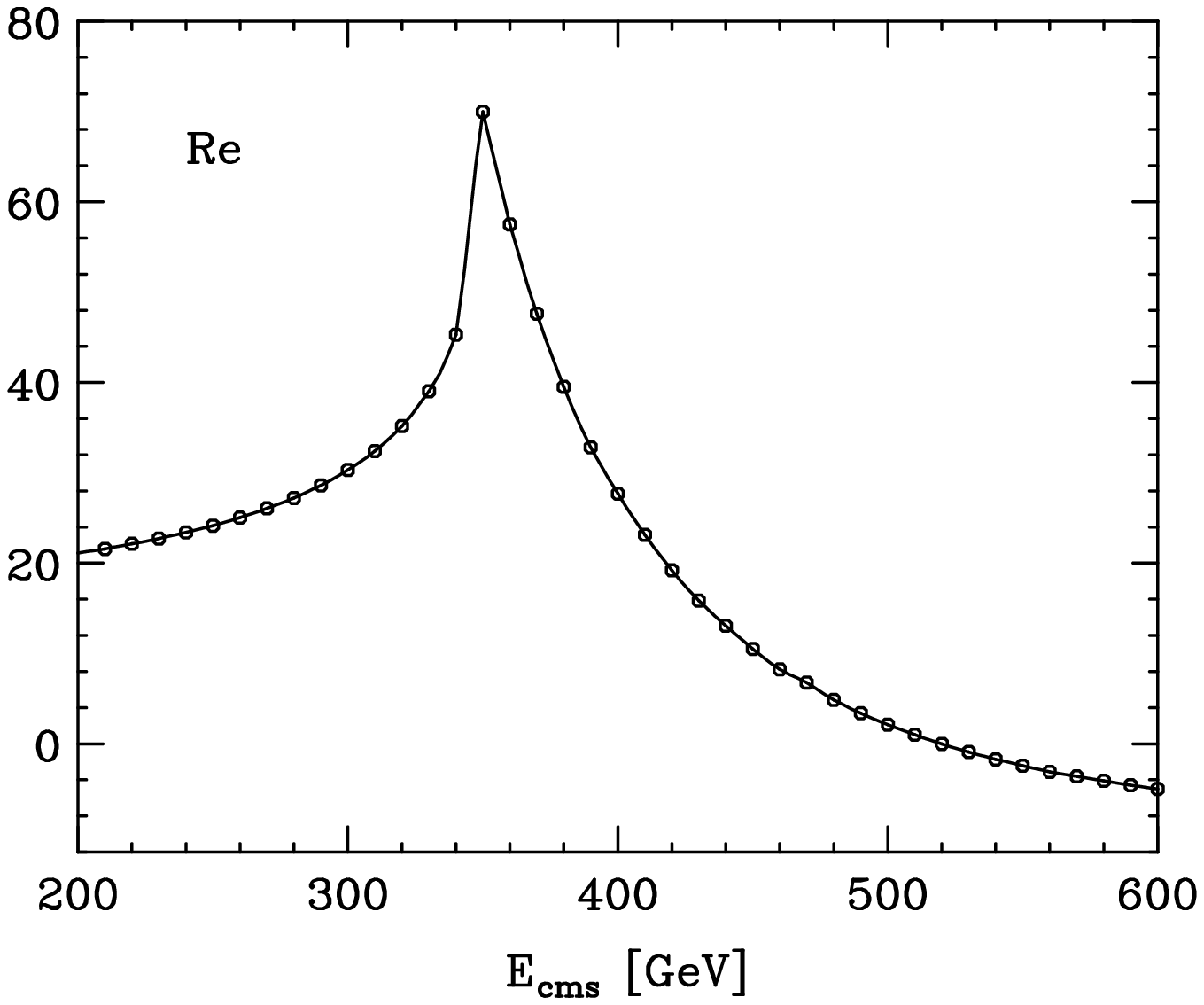}} 
\put(10,-10){\includegraphics[width=5.cm, angle=90]{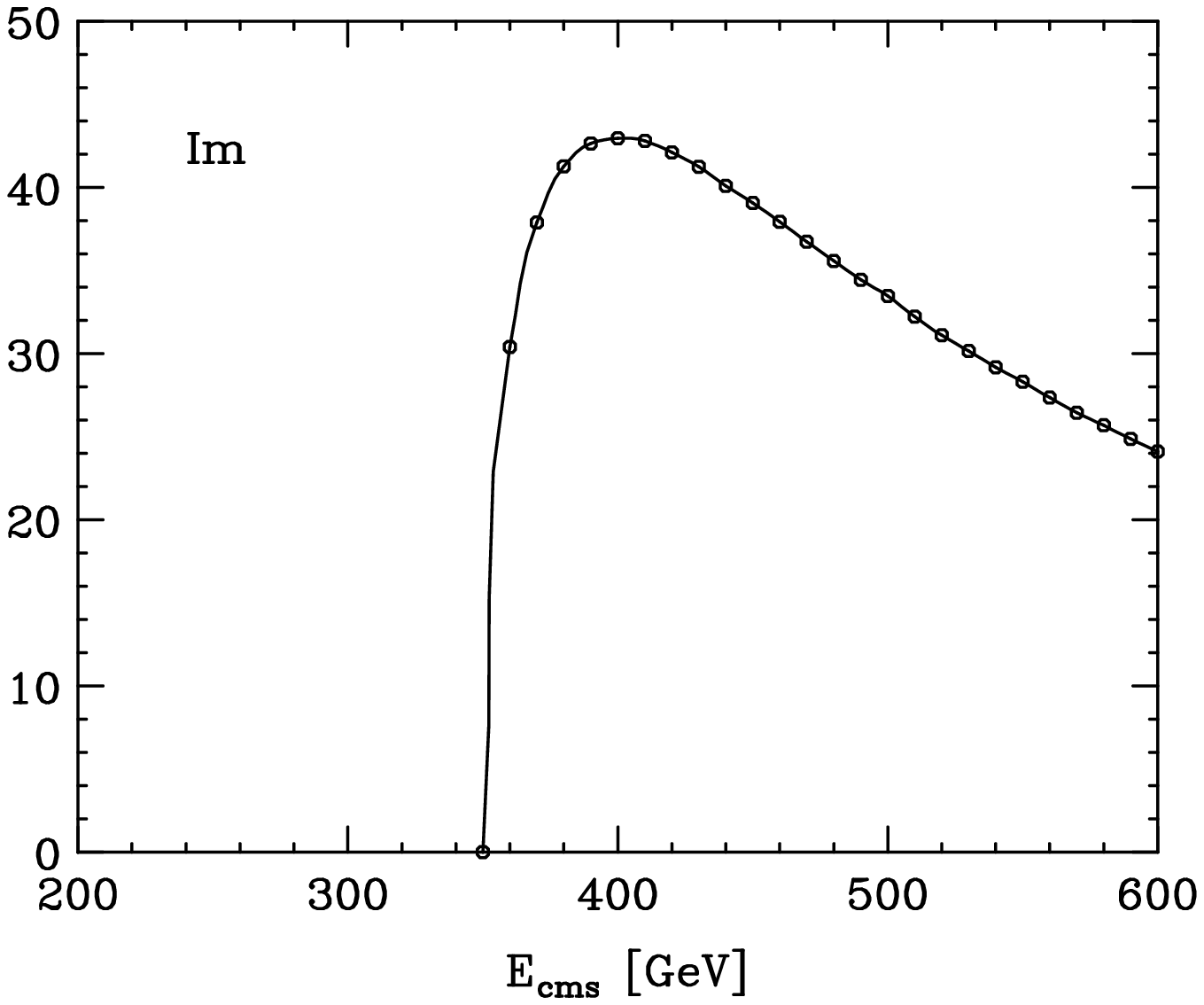}}
\end{picture}
\caption{Scan of the $2m_t = 350$ GeV threshold of the 4-dimensional scalar hexagon function 
which corresponds kinematically to the 
right Feynman diagram of Fig.~\protect\ref{FigHex}.}
\label{hexagon-scan}
\end{figure}
For details of the integration methods see \cite{Binoth:2002xh,Kauer:2002hk}.

\section{CONCLUSION}

To make reliable phenomenological studies for collider experiments operating at the TeV scale 
1-loop calculations with many external particles are mandatory. 
In this talk I have outlined recent developments concerning 
the analytic and numeric evaluation of 1-loop Feynman diagrams.
Using reduction methods a compact result for the 3-gluon 2-photon amplitude
was presented. Concerning numerical methods we have developed an approach
to successfully integrate hexagon functions numerically. Merging 
and applying these techniques to more challenging situations is presently under study.  

\section*{ACKNOWLEDGMENTS}

I would like to thank the conference organizers for the stimulating
conference in Zinnowitz. 

\bibliography{bibli_nlo} 

\end{document}